\documentclass[12pt]{article}
\newcommand{\sh}[1]{#1 \!\!\!/}

\begin{document}
\begin{center}
\begin{large}
{\large On the twist-3 contribution to $h_L$ in
the instanton vacuum} \\[0.5cm]
\end{large}
{\bf B.\ Dressler}$^{\rm a}$,
{\bf M.V.\ Polyakov}$^{\rm a,b}$,
\\[0.5cm]
$^{a}${\em Institut f\"ur Theoretische Physik II,
Ruhr--Universit\"at Bochum, \\ D--44780 Bochum, Germany}
\\
$^{b}${\em Petersburg Nuclear Physics Institute, Gatchina, \\
St.Petersburg 188350, Russia}
\end{center}

\begin{abstract}
\noindent
We show that the instanton model of the QCD vacuum indicates
the parametric smallness of the twist-3 contributions to the
polarized structure function $h_L$. This smallness is related
to the diluteness of the QCD instanton vacuum.
\end{abstract}

\noindent
{\bf1.} Among higher twist parton distributions the twist-3 distributions play a
special role. In many physical observables the twist-3 distributions
enter not suppressed by powers of the hard scale relative to twist-2
distributions. Therefore the determination of twist-3 distributions
does not encounter the
conceptual problems of the separation of power suppressed
contributions from those
that are suppressed by only logarithms.
Examples of such observables are spin asymmetries in DIS on
transversely polarized targets ($g_2$)
\cite{AnselminoEfremovLeader}
and single spin azimuthal asymmetries in semi-inclusive production of
hadrons ($h_L$) \cite{Collins,Mulders,Kotzinian}. The experimental
data of DIS
on transversely polarized targets have already reached the precision to
estimate the twist-3 contributions to the observables,
see recent measurements by E155 collaboration \cite{E155}.
Recent HERMES and SMC data on single spin azimuthal asymmetries
\cite{HERMES,SMC} provide the possibility to estimate the quark
transversity distribution $h_1$ in the nucleon if, among other things,
one would be able to estimate the twist-3 contribution to $h_L$.
The objective of this report is to make an estimate of the size of the twist-3
contribution to $h_L$ in the instanton model of the QCD vacuum.

The twist-3 distributions are given by nucleon matrix elements of
mixed quark-gluon operators. These matrix elements are very sensitive to the
correlations of non-perturbative gluon and quark fluctuations in the QCD
vacuum. The theory of such fluctuations is provided by the instanton
model of the QCD vacuum \cite{DP86} (for a review see
\cite{D96_Varenna,SchSh96}).
A nice feature of the instanton model of the QCD vacuum is the
existence of a small parameter--the ratio of the average instanton
size $\bar \rho$ to the average distance between instantons $\bar R$
($\bar \rho/\bar R\approx 1/3$). This parameter was first anticipated
in ref.~\cite{Shur82} from phenomenological considerations, obtained
in dynamical calculations of \cite{DP84} and recently confirmed by
direct measurements on the lattice \cite{latt}.

In ref.~\cite{DPW96} a method was developed to calculate hadronic matrix
elements of mixed quark gluon operators in the instanton vacuum.
Later this method was applied to estimates of higher twist operators
\cite{BPW97}. In particular it was shown that the twist-3 contribution
to the structure function $g_2$ is parametrically small relative to twist-2
and twist-4 contributions. For example, the third moment of $g_2$
\begin{equation}
\int_0^1 dx \, x^2 g_2 (x, Q^2 )
\;\; = \;\; -\frac{1}{3} a^{(2)}
\; + \; \frac{1}{3} d^{(2)}
\; + \; O\left(\frac{1}{Q^2}\right)
\label{g2}
\end{equation}
can be splitted into the twist-2 part $a^{(2)}$ and twist-3 part $d^{(2)}$.
In the instanton vacuum,
the twist-2 part is
parametrically of order
\begin{equation}
a^{(2)}\sim (\bar \rho^2/\bar R^2)^0\sim 1
\end{equation}
whereas the twist-3 part behaves like
\begin{equation}
d^{(2)}\sim (\bar \rho^2/\bar R^2)^2\log(\bar \rho^2/\bar R^2)\sim 10^{-3}
\label{tw3ofg2}
\end{equation}
(see \cite{BPW97} for details). This strong suppression of the twist-3
contribution relative to twist-2 one
is related to the specific spin-colour
structure of the instanton field and its  properties under
conformal transformations. Using this fact one can conclude that
the suppression of the twist-3 part persists also for higher moments,
not only for the lowest one. Here we repeat the analysis for the
lowest moment
of $h_L^{tw3}$, leaving the general proof for a comprehensive paper
\cite{inprep}.
\vspace{0.3cm}

\noindent
{\bf2.}
The Mellin moments of $h_L(x)$ can be splitted into twist-2 and
twist-3 part \cite{hl}
\begin{equation}
\mathcal{M}_{n}[h_{L}]\equiv \int_0^1 dx\ x^n\ h_L(x)=
\frac{2}{n+2} \mathcal{M}_{n}[h_1]+ \mathcal{M}_{n}[h_{L}^{tw3}]\, ,
\end{equation}
where the first term is related to the Mellin moment of the twist-2
transversity quark distribution $h_1(x)$
\cite{hl1}. The moments of \( h_{L}^{tw3} \)
are related to the following matrix elements of mixed quark-gluon operators
\cite{hl}:
\begin{equation}
\mathcal{M}_{n}[h_{L}^{tw3}]=
\sum _{l=2}^{\frac{n+1}{2}}\left( 1-\frac{2l}{n+2}\right) b_{nl}(\mu ^{2})
\end{equation}
with
\begin{equation}
\langle pS|R_{nl}^{\mu _{1}\dots \mu _{n}}(\mu ^{2})
|pS\rangle =2b_{nl}(\mu ^{2})M^2(S^{\mu _{1}}
P^{\mu _{2}}\dots P^{\mu _{n}}-\mathrm{traces})
\end{equation}
The general form of the operators $R_{nl}^{\mu _{1}\dots \mu _{n}}(\mu ^{2})$
can be found in \cite{hl}.

Here we shall be interested in the lowest
non-vanishing moment $n=3$
\begin{equation}
\mathcal{M}_{3}[h_{L}^{3}]=\frac{1}{5}b_{32}(\mu ^{2})
\end{equation}
with $b_{32}(\mu ^{2})$ defined through the matrix element
\begin{equation}
\langle PS|R_{32}^{\delta \alpha \beta }(\mu ^{2})|PS\rangle =
2b_{32}(\mu ^{2})M^{2} \mathcal {S}(S^{\delta }P^{\alpha }P^{\beta }-
\mathrm{traces}) \;.
\label{matrel}
\end{equation}
where $\mathcal{S}$  denotes the symmetrization of Lorentz indicies and
the local operator has the form
\begin{equation}
R_{32}^{\delta \alpha \beta }=\frac{1}{2}\mathcal{S}\bar{\psi }\sigma ^{\gamma
  \delta }i\gamma _{5}[i\nabla ^{\alpha },iF^{\beta }_{\: \: \gamma }]\psi
-\mathrm{traces}
\end{equation}
or equivalently
\begin{equation}
R_{32}^{\delta \alpha \beta }=
-\frac{i}{2}\mathcal{S}\bar{\psi }\sigma^{\gamma \delta }
\gamma _{5}(D^{\alpha ac}F^{\beta \: c}_{\: \: \gamma })
\frac{\lambda ^{a}}{2}\psi -\mathrm{traces} \;.
\label{operator}
\end{equation}
We shall compute the matrix element (\ref{matrel}) in the instanton
model of the QCD vacuum using the technique of refs.~\cite{DPW96,BPW97}.

The effective low--energy theory one derives
from the instanton vacuum is formulated in terms of degrees of freedom
which are pions (Goldstone bosons) and massive ``constituent'' quarks. It
is described by the effective action \cite{DP86,D96_Varenna}
\begin{equation}
S_{\rm eff} \;\; = \;\; \int d^4 x \;
\bar\psi (x) \left[
i \gamma^\mu \partial_\mu \; - \;
M F(\stackrel{\leftarrow}{\partial} ) e^{i \gamma_5 \tau^a \pi^a (x)}
F(\stackrel{\rightarrow}{\partial} ) \right] \psi (x) .
\label{S}
\end{equation}
Here, $M$ is the dynamical quark mass generated by the spontaneous breaking
of chiral symmetry; parametrically it is of order:
\begin{eqnarray}
M\bar\rho &\sim & \left(\frac{\bar\rho}{\bar R}\right)^2 ,
\end{eqnarray}
and $F(k)$ is a form factor proportional to the wave function of the
instanton zero mode, which drops to zero for momenta of order
$k \sim \bar\rho^{-1}$. Mesonic correlation functions computed either
with the effective action, Eq.(\ref{S}), using the $1/N_c$--expansion
\cite{DP86} or by more elaborate numerical simulations \cite{SchSh96}
show excellent agreement with phenomenology.

In order to find the parametric behaviour of $b_{32}$ in the packing
fraction $\bar\rho^2/\bar R^2$ it is enough to compute the matrix
element (\ref{matrel}) in constituent quark states. In order to
accomplish this one has to transform the operator (\ref{operator}) into
the corresponding effective operator in the effective low-energy theory
(\ref{S}). The details of such transformation can be found in
\cite{DPW96,BPW97}. Here we only report the main technical steps.

First we compute  the covariant derivative of the gluon field strength
on the field of one instanton (anti-instanton) $I(\bar{I})$:
\begin{eqnarray}
\lefteqn{ D_{\alpha }^{ac}F_{\beta \gamma }^{c}(x)_{I(\bar{I})} } \nonumber \\
& = &
(\eta ^{\mp })^{a}_{\lambda \rho }
\frac{-48\bar\rho ^{2}}{(x^{2}+\bar\rho
  ^{2})^{3}} \bigg[ \bigg(\frac{x_{\alpha }x_{\beta } x_{\lambda
    }}{x^{2}}-\frac{1}{6}(x_{\alpha } \delta _{\beta \lambda }+x_{\beta }
\delta _{\alpha \lambda }+x_{\lambda } \delta _{\alpha \beta })\bigg )
\delta _{\gamma \rho }-(\beta \leftrightarrow \gamma )\bigg] \,.
\end{eqnarray}
Its Fourier transform has the form
\begin{equation}
\mathcal{K}_{\lambda \rho \alpha \beta \gamma }(k)=
i\bar{\rho }^{2} \mathcal{K}(k^{2})
\bigg[ \bigg(\frac{k_{\alpha }k_{\beta }
k_{\lambda }}{k^{2}}-\frac{1}{6}
(k_{\alpha }\delta _{\beta \lambda }+
k_{\beta }\delta _{\alpha \lambda }+
k_{\lambda }\delta _{\alpha \beta })\bigg )
\delta _{\gamma \rho }-
(\beta \leftrightarrow \gamma )\bigg] \, ,
\end{equation}
where
\begin{equation}
\mathcal{K}(k^{2})=
(24\pi )^{2}\bigg
[-\frac{16}{t^{6}}+
\bigg (\frac{16}{t^{5}}+\frac{4}{t^{3}}+\frac{1}{4t}\bigg )
K_{1}(t)+\bigg (\frac{8}{t^{4}}+\frac{1}{t^{2}}+\frac{1}{24}\bigg )
K_{0}(t) \bigg], \ t=|k|\bar\rho \,.
\end{equation}
Here $K_{\nu}(t)$ are modified Bessel functions.
Using this result we can easily derive the form of the effective
operator\footnote{We give the expression for the case of one
flavour which is enough to compute the matrix element of the
operator between constituent quark states}:
\begin{eqnarray}
R_{\delta \alpha \beta }^{\mathrm{eff}}(x) & = &
  \frac{M}{N_{c}}\int \mathrm{d}^{4}z \mathcal{K}_{\lambda \rho
  \alpha \beta \gamma }(x-z) \nonumber \\
& & \times \psi ^{\dagger }(x)\sigma _{\gamma \delta } \gamma
  _{5}\frac{\lambda ^{a}}{2}\psi (x) \psi ^{\dagger }(z)\frac{\lambda ^{a}}{2}
\sigma _{\lambda \rho }\frac{1\pm \gamma _{5}}{2}\psi (z) \, .
\end{eqnarray}
Now it is easy to compute the matrix element of this operator
in constituent quark states
\begin{eqnarray}
\lefteqn{ \langle p|R_{\delta \alpha \beta }^{\rm {eff}}|p\rangle  =
 \frac{iM}{2}\int \frac{\mathrm{d}^4 k}{(2\pi)^4} \mathcal{K}_{\lambda \rho
 \alpha \beta \gamma }(k) } \nonumber \\
 &  & \times \bigg [\frac{F(p)F(p-k)}{(p-k)^{2}+M^{2}F^{4}(p-k)}
 \mathrm{Tr}\Big [\Lambda _{p,S}\sigma _{\gamma \delta }\gamma _{5}\Big (\sh
 {p}-\sh {k}+iMF^{2}(p-k)\Big )\sigma _{\lambda \rho }\frac{1\pm \gamma
 _{5}}{2}\Big ] \nonumber \\
 &  & \hspace {0.22cm}+\frac{F(p)F(p+k)}{(p+k)^{2}+M^{2}F^{4}(p+k)}
 \mathrm{Tr}\Big [\Lambda _{p,S}\sigma _{\lambda \rho }\frac{1\pm \gamma
 _{5}}{2}\Big (\sh {p}+\sh {k}+iMF^{2}(p+k)\Big )\sigma _{\gamma \delta
 }\gamma _{5}\Big ]\bigg ] \, ,
\end{eqnarray}
where the projector on quark states with definite momentum and polarization
vector has the form
\begin{equation}
\Lambda _{p,S} = u(p,S)\bar{u}(p,S) = \frac{-i\sh {p}+M}{2} \Big(1+i
\gamma_{5}\sh {S}\Big) \, .
\end{equation}
The traceless part of the operator
$R_{\delta \alpha \beta }^{\mathrm{eff}}$ which is related to
$b_{32}$ can be isolated by contracting the Lorentz indices
$\delta \alpha \beta$ with a light-cone vector $n$, such that
$n\cdot p$ and $n\cdot S$ are non-zero:
\begin{equation}
n_{\alpha }n_{\beta }n_{\delta }\langle p|
R_{\delta \alpha \beta }^{\mathrm{eff}}|p\rangle =
2M^{2}I(p)(n\cdot p)^{2}(n\cdot S) \, .
\end{equation}
The quantity \( b_{32} \) is related to $I(p)$ as
$b_{32}=-I(M)$.
The expression for $I(p)$ is given by a simple integral:
\begin{equation}
I(p)=
\bar{\rho }^{2} \int \frac{\mathrm{d}^4 k}{(2\pi)^4}
\frac{F(p)F^{3}(p-k) \mathcal{K}(k^{2})}{(p-k)^{2}+
M^{2}F^{4}(p-k)}\bigg [\frac{k\cdot p}{p^{2}}-
2\frac{(k\cdot p)^{3}}{k^{2}p^{4}}\bigg ] \, .
\label{iofp}
\end{equation}
Its small $p$-behaviour has obviously the form
\footnote{The integral (\ref{iofp}) has generic form (for $n=3$):
$$
I_n(p)\propto\bar{\rho }^{2} \int \frac{\mathrm{d}^4 k}{(2\pi)^4}
\frac{F(p)F^{3}(p-k) \mathcal{K}(k^{2})}{[(p-k)^{2}+
M^{2}F^{4}(p-k)] k^2} C_n^{(1)}\biggl( \frac{k\cdot p}{|k||p|}
\biggr)
|k|^n |p|^{2-n}\sim p^2\bar\rho^2 \log(p\bar\rho) \, .
$$ where $C_n^{(1)}$ is Gegenbauer polynomial}:
\begin{equation}
I(p)\sim p^2 \bar{\rho }^{2}\ln (|p|\bar{\rho }) \, .
\end{equation}
{}From this we conclude that $b_{32}$ is parametrically suppressed
by the packing fraction of the instanton liquid
\begin{equation}
b_{23}\sim (\bar \rho^2/\bar R^2)^2\log(\bar \rho^2/\bar R^2) \, ,
\end{equation}
i.e. as for twist-3 contribution to $g_2(x)$, see
eq.~(\ref{tw3ofg2}). This is the main result of this report.
We can expect that the twist-3 part of $h_L$ is also numerically
much smaller than its twist-2 part because the twist-2 part of $h_L$
behaves in the packing fraction as
$\sim (\bar \rho^2/\bar R^2)^0$ \cite{PP96}.
The obtained suppression of the twist-3 part of $h_L$ refers to
a low normalization point of order $\sim 1/\bar\rho\approx 0.6$~GeV. Under
evolution to higher normalization points the twist-3 part of $h_L^{tw3}$ dies
out faster than $h_L^{tw2}$ \cite{hlevol}, so that the suppression of
$h_L^{tw3}$ relative to $h_L^{tw2}$ will be even more pronounced at higher
$Q^2$.

Numerically one gets $b_{32}=-I(M)=-0.014$ at $M\bar{\rho }=0.58$. From this
we can make a rough estimate of the ratio

\begin{equation}
\frac{\mathcal{M}_{3}[h_{L}^{tw3}]}{\mathcal{M}_{3}[h_{L}^{tw2}]}
\sim  10^{-2}\, .
\end{equation}
Let us note that in the bag model the corresponding ratio is about 10
times larger \cite{hl,bag}.

\vspace{0.3cm}

\noindent
{\bf 3.}
To summarize, we have shown that the instanton vacuum with its inherent
small parameter, $\bar\rho / \bar R$, implies a parametrical (and
numerical) hierarchy of the spin--dependent
twist--2 and --3 matrix elements: $h_L^{tw3}\ll h_L^{tw2}$.
The same hierarchy was observed for twist--2 and --3  contributions
to $g_2$ \cite{BPW97}, which seems to be confirmed by recent measurements by
E155 collaboration \cite{E155}.

\medskip
{\footnotesize We would like to thank A.V. Efremov,
K.~Goeke, A.Kotzinian, P.V. Pobylitsa and C.~Weiss
for fruitfull discussions.}

\end{document}